\documentclass[acmlarge,screen,nonacm]{acmart}
\renewcommand\footnotetextcopyrightpermission[1]{} % removes footnote with conference information in first column
\AtBeginDocument{%
  \providecommand\BibTeX{{%
    \normalfont B\kern-0.5em{\scshape i\kern-0.25em b}\kern-0.8em\TeX}}}

\setcopyright{acmcopyright}
\copyrightyear{2018}
\acmYear{2018}
\acmDOI{XXXXXXX.XXXXXXX}

\begin{document}

\title{A Retrospective on ICSE 2022}

\author{Cailin Winston, Caleb Winston, Chloe Winston, Claris Winston, Cleah Winston}
\authornote{All authors contributed equally to this document.}
\email{cailinw@cs.washington.edu}
\affiliation{%
  \institution{Paul G. Allen School of Computer Science, University of Washington, Seattle}
  \city{Seattle}
  \country{USA}
}

\maketitle

\section{Introduction}
The 44th International Conference on Software Engineering (ICSE 2022) was held in person from May 22 to May 27, 2022 in Pittsburgh, PA, USA. Since ICSE was held as a solely virtual conference for the last two years, the opportunity to interact with other members of the software engineering community in person and to engage in insightful discussions in a physical room was greatly welcomed.

Each day was organized into paper sessions, poster sessions, and Birds of a Feather (BoF) sessions, in addition to plenty of time for networking. Each paper session consisted of around 6-10 5 minute talks and time for questions for the authors. The Birds of a Feather sessions allowed for a broader discussion on a topic; the sessions varied in terms of topics and format.

In this document, we summarize themes of research that we observed at the conference. We organized these under the following topics (in no particular order):

\begin{enumerate}
    \item Autonomous Vehicles (\autoref{sec:avs})
    \item API Misuse (\autoref{sec:misuse})
    \item Machine Learning Libraries (\autoref{sec:mllibs})
    \item Machine Learning Automating Software Engineering (\autoref{sec:mlautose})
    \item Communities (\autoref{sec:communities})
    \item Education (\autoref{sec:education})
\end{enumerate}

\section{Autonomous Vehicles} \label{sec:avs}

With companies such as Tesla, Waymo, and NVIDIA making advances in autonomous vehicle technology, it is important for the software engineering community to discuss the current pressing challenges in developing and testing these hardware-software systems and improve the testing methods. Autonomous vehicles are highly safety-critical technologies, and thus reliability must be ensured.

At ICSE 2022, the most pressing challenges were discussed. Issues around who/what is liable for bugs and skill decrement of future drivers were mentioned broadly, but these remain open issues to be tackled in a different forum. The issues of little effective regulation were quickly laid aside, and the discussion was focused on the need for a benchmark for safety/reliability and the challenges of testing these safety-critical systems. A benchmark for safety and reliability is lacking because it is hard to define what is "sufficient dependability" for autonomous vehicles, and it is hard to test a system to prove that reliability criteria are satisfied. Furthermore, different definitions of safety might exist. Hu et al. \cite{hu2022if} addressed this challenge by defining reliability requirements for machine vision components (MVCs), such as those critical in autonomous vehicle systems, based on the baseline of human performance and by developing methods for specifying correctness and for checking whether requirements are satisfied. These requirements ensure that the performance of an MVC should not be affected by image transformations that would not affect a human’s vision and judgment.

Furthermore, software is oftentimes the easiest component to change in hardware-software systems, so it sometimes becomes a catch-all for bugs that arise, and thus the need to address the challenges in testing these complex systems is important. Woodlief et al. \cite{woodlief2022semantic} presented a mutation-based approach to test machine vision components that are used in autonomous vehicles by generating mutations based on semantics from an image dataset. Regarding testing the entire system, simulators exist, but methods for testing the simulators have not been developed. However, the priority for the large companies that are developing these technologies is maximizing revenue, which may not always align with the ideal criteria for safety.

\section{Fighting Misuse}  \label{sec:misuse}

Research presented at this year’s ICSE on API misuse detection addressed one key issue at large: false positives in mined instances of API misuse. API misuse occurs when an API has some inner workings that is misunderstood by the user of the API and this results in code that may result in an error. API misuse detection relies on a large set of mined instances of API misuse. Unfortunately, as Lamothe et al. discovered, mined instances of API misuse can turn out to be mostly false positives \cite{lamothe2021assisting}.

Lamothe et al. addressed this problem of false positives with a data augmentation strategy by automatically producing complementary examples of API usage. Kang et al. addressed this problem with an active learning strategy where different API usage examples are selected for manual inspection based on their informativeness \cite{kang2021active}. Since API misuse can lead to vulnerabilities or failures, it is exciting to see work to more effectively mine instances of API misuse.

\section{Machine Learning Libraries}  \label{sec:mllibs}

While there has been a larger trend of machine learning (ML) with and for software engineering (SE), several papers tackled the more specific issue of testing and debugging ML libraries. While some past approaches for detecting bugs in deep learning libraries use differential testing (i.e., testing whether similar functions have the same functionality between different libraries), a noted issue with this is the reliance on multiple libraries for the same set of functionality. As an example, both Pytorch and Tensorflow and popular deep learning libraries that provide similar functionality but with different APIs. Several within-library differential testing techniques were suggested. EAGLE \cite{wang2022eagle} defines equivalence rules and develops equivalent graphs that should yield the same results. The functionality of these equivalent graphs is verified to be identical. FreeFuzz takes the approach of mining usage of functions from open source code and carries out differential testing between similar concepts (e.g., CPU vs GPU computation should yield the same results) \cite{wei2022free}. EAGLE and FreeFuzz are able to develop bugs, but another issue with debugging machine learning libraries is that these are frequently updated, and thus, the expected functionality of various methods may change over time. DiffWatch \cite{prochnow2022diffwatch} addresses this issue by identifying differential tests and automatically detecting changes in deep learning libraries that might affect the expected outcome of the tests. This can prompt developers to update their tests accordingly.

DeepStability tackles a subtler type of bugs - those having to do with mathematical instability. Kloberdanz et al. studied a number of commit messages in Pytorch that indicated a bug related to numerical instability \cite{kloberdanz2022deepstability}. They manually verified what type of commit it was, whether it was related to numerical instability, and whether it was a new unit test or a fix to a previous bug. The authors characterized the root causes and the patches for the discovered issues and used this information to identify numerical instability in other functions such as cosine similarity.

Another set of research addressed the issue of debugging the use of deep learning libraries. Keeper \cite{wan2022automated} is a tool for testing software using cognitive ML APIs that utilizes pseudo-inverse functions, symbolic execution, and automatic generation of test inputs to fulfill branch coverage and identify root causes of bugs. DeepDiagnosis \cite{wardat2021deepdiagnosis} is an approach for debugging DNN programs by identifying and localizing faults, such as an exploding tensor or loss not decreasing during training, as well as suggesting fixes for the root causes of the faults. Another work \cite{tizpaz2022fairness} proposes enabling ML libraries to search for hyperparameter configurations that encourage learning a fair model, given the dataset.

\section{ML is Automating SE}  \label{sec:mlautose}

The automation of various software engineering (SE) tasks by using
machine learning (ML) has been a research trend for many years.
At ICSE 2022, we observed specific themes of work around
explainability, variables, and automatic program repair.

\subsection{Explainable Models of Code}

One topic that came up multiple times was the topic of explainability. The idea was often that if we really want to use ML for SE applications, we need to be able to explain to humans why an ML model's prediction about some code should be trusted.

Regarding ML models for program repair, Noller et al. surveyed developers and discovered that not only do most current program repair tools not produce high-quality patches but also developers do not seem interested in human-in-the-loop interaction to collaboratively develop better patches \cite{noller2022trust}. Regarding summarizing code diffs, Cito et al. develop a method for producing counterfactual explanations for models of code \cite{cito2021counterfactual}. The focus of this work is models for automated code review that predict some metric such as performance given a code diff. In automated code review, counterfactual explanations are critical because humans would want to understand what aspect of the diff causes a predicted performance regression for example.

\subsection{Variables}

There was also quite a bit of research centered on leveraging variables and their names to develop effective models of code. Yang et al. focused on adversarially training models of code by changing variable names \cite{yang2022natural}. Variable names are changed by greedily selecting more tokens that were more influential for correct predictions. Chen et al. focused more specifically on pretraining the representations of individual variable names \cite{chen2021varclr}. They determined similarity of variable names and apply contrastive loss to learn a more useful representation of variable names. Finally, Patra et al. presented work on using a neural classification model to classify learned representations of variable names and values as consistent or inconsistent \cite{patra2021nalin}.

\subsection{Automatic Program Repair}

There have been multiple successful approaches to automatic program repair (APR). Given this, Noller et al. interviewed software practitioners to determine the usability and acceptability of current APR approaches \cite{noller2022trust}. The results of the survey indicated several issues with current APR approaches, listed below:
\begin{itemize}
\item Finding patches automatically can be time-consuming. Software practitioners prefer patches to be found within 2 hours.
\item Software practitioners prefer fewer than five patches to be generated to reduce the amount that needs to be reviewed.
\item While software practitioners generally want some interaction, they prefer low-interaction methods.
\item APR methods may generate patches that overfit to a specific test suite.
\item The inability to explain patches makes software practitioners reluctant to trust them. Additional tests generated by APR methods would help practitioners trust the accuracy of the patch.
\end{itemize}

These findings will help direct future research and development of APR techniques, and several other papers presented at the conference were directly targeted towards these issues. PropR incorporates property-based testing into APR. Property-based tests are tests defined by properties of input-output pairs \cite{gissurarson2022propr}. Thus, an infinite number of tests can theoretically be produced, where the properties of the output are verified. PropR is implemented in Haskell and creates patches faster, and importantly patches that do not overfit due to the additional property testing.

PropR utilizes underlying semantics-based repair techniques. However, deep learning approaches have been explored too. The complexity of these approaches though present several challenges, out of which two were directly addressed by papers this year.
\begin{itemize}
\item Deep learning approaches to this point have only been capable of generating single-line patches. DEAR, however, uses a divide-and-conquer approach to successfully create multi-line and multi-hunk patches \cite{li2022dear}.
\item Deep learning approaches require a loss function to minimize. The typical loss function used for APR techniques is based on syntactical differences between generated and “ground-truth” patches. This means that patches that do not even compile may have a lower computed loss than slightly incorrect patches that do compile. This is likely one of the reasons behind the low-quality patches generated through deep learning approaches. RewardRepair developed a new loss function that executes predicted patches and penalizes patches that do not compile and thereby enables improvement over state-of-the-art approaches \cite{ye2021neural}.
\end{itemize}

These results are promising; however, a discussion with the authors of the above papers revealed that there are a number of uncertainties in the future of deep learning-based APR methods.
\begin{itemize}
\item Deep learning is fundamentally data-driven. Thus, large quantities of data are needed to train these models, and this is difficult to obtain because it requires large quantities of bugs and associated patches. This being said, the ideas of property testing as used in PropR can be applied to augment training datasets.
\item As mentioned above, deep learning approaches so far do not perform well on APR tasks, yet DEAR and RewardRepair have taken steps towards tackling these issues.
\item There is also a concern about the sustainability of deep learning in general due to the resource consumption required for training models. This is an active area of research though in deep learning.
\item Deep learning approaches may also learn a limited type of patch based on the types of patches in the training dataset. For this issue as well, the ideas behind PropR may be applicable.
\end{itemize}
Overall, it is possible that a combination of symbolic, semantics, and deep learning based approaches may prove effective. For example, deep learning approaches could be used to repair programs viewed from the semantics, whereas program synthesis approaches are used to generate code from the repaired semantics.

\section{Software Engineering Communities} \label{sec:communities}

A theme of research presented at ICSE 2022 was studying the communities surrounding software engineering. Examples of communities included GitHub, StackOverflow, and Twitter.

\subsection{Open Source Software (GitHub)}

GitHub represents some of the largest software engineering communities. Given its open nature, it has proven particularly accessible to researchers to investigate SE communities.

It’s important to foster a safe and positive environment in open source GitHub projects. Miller et al. and Qiu et al. explored the \textbf{toxicity} that often surrounds open source communities such as on GitHub \cite{miller2022did,qiu2022detecting}. The authors have described the attitude of unreasonably high expectations towards open source developers, and analyzed toxic GitHub issue discussions and interpersonal conflict in code review which result from this. The toxicity that exists in OSS is shown to be different from the toxicity that exists in other social media platforms such as Reddit, so doing more analysis and finding ways to detect toxic behavior in OSS, is a first step to being able to mitigate this attitude.

On a similar note, positively \textbf{welcoming newcomers} into such OSS projects is very important. Guizani et al has worked on creating a maintainer dashboard that works to attract and retain new contributors by highlighting achievements of contributors encouraging maintainers to recognize new contributors \cite{guizani2022attracting}. Currently, many GitHub repositories have bots for pull requests. Russel et al. discussed the noisiness of these kinds of bots, and how they can be overwhelming for newcomers to GitHub repositories - instead they suggested creating better ways to represent the information of bots, such as clearer  summaries \cite{wessel2022bots}.

Another aspect that was discussed is \textbf{gender diversity in OSS}. Studies demonstrated the current gender imbalance in open source contributions. Rossi et al. and Prana et al. conducted a large global study of how the gender gap of contributions in OSS has changed over the years, and in different worldwide regions \cite{prana2021including,rossi2022worldwide}. They found that over the last 20 years, the gender gap has been gradually decreasing worldwide. However, over the COVID-19 pandemic, while the contributions by male participants has not changed much, there was a significant decrease in the contributions of female participants to OSS, increasing the gender gap. Again, these patterns were consistent across worldwide regions. While the general trends were similar, they found that certain regions such as regions in Africa have been experiencing a growth in gender diversity faster than others. They noted while there has been much positive change in the last decade or two, there is still much work to be done to reduce the gender gap in OSS.

Much of this research leverages open-source repositories but doesn’t venture into closed-source or internal software. A question was raised to the above authors in Q\&A following their presentations about how their research might translate to closed-source projects. Separately, Qiu et al. compared GitHub open-source projects and Google internal code review to determine how detecting toxic behavior varies between open- and closed-source \cite{qiu2022detecting}. They found that both have platform specific features so you can’t use the exact same model when finding toxic comments in open source projects or pushback in internal code review. 

\subsection{Platforms for Discussion (StackOverflow and Twitter)}
StackOverflow is a popular resource for programmers, and there were a couple papers that delved into how StackOverflow is used. Some research discussed how expert programmers use stack overflow in a different and usually more effective way than novices - based on this, they suggested that it may be worthwhile to consider teaching students how to properly look for online help on sites such as stack overflow.  Somewhat relatedly to StackOverflow, Xu et al. proposed Post2Vec which leveraged tags and other metadata to learn better representations of StackOverflow \cite{xu2021post2vec}. This work could also help developers more easily discover posts on StackOverflow.

\subsection{Remote Work}
Another theme of discussion was how to move to and improve remote work, emphasizing that having a positive setup to help with accessibility in case some people need to work remotely. Research studied how remote work affected people’s productivity - the same aspects of remote work that are advantageous to some people, such as schedule flexibility, may be disadvantageous to others. Some suggestions to help ease the challenges that come with remote work were creating a 10 minute rule time limit for meeting (starting meetings 10 minutes after the hour and ending 10 minutes before to thesaurus to give everyone at least a 20 minute break between meetings, and help with productivity during the meeting), making sure to end virtual meetings on time, and chatting about personal things virtually to help maintain connection. Another challenge discussed by Santos et al., is coordination in remote environments - they discussed how coordination depends on cohesion \cite{santos2022grounded}. 

\section{Education} \label{sec:education}

Instructors from different universities share techniques that they have used to teach software engineering and design. Many papers at ICSE 2022 emphasized the importance of bringing software engineering into the classroom and described techniques that can be used when teaching software engineering and design. 
\subsection{Techniques for Teaching Software Engineering}
Through various papers and discussion sessions, many points were brought up about techniques to teach software engineering and design. 

\begin{itemize}
\item \textbf{Peer feedback.} Providing peer feedback to students can assist them in learning software design. As Rukmono et al. discusses, students appreciate both positive feedback and critical feedback that is specific with examples \cite{rukmono2022guiding}. They also learned that students should be taught how to provide useful peer review to other student’s projects and that students who are more engaged with the course material are more likely to implement feedback from peer review.
\item \textbf{Enforcing certain guiding principles in design.} One professor suggested thinking about the inclusivity, security, and fairness of the user as a guiding principle. Another suggestion was to encourage students to think more about users when designing software, and to carefully consider any assumptions made about them. Students should be encouraged to think about the end user of the software, as well as maintain code that is readable by everyone.  
\item Some professors discussed an assignment, where students would look at a \textbf{real system} that is currently used, and document its software design. While it can be challenging to find the necessary information to complete such a project, the professors found that it benefited the students to see how software engineering is applied in the real world.  
\item \textbf{Hackathons.} Hackathons could also be useful in teaching software engineering. Affia et al. suggests introducing hackathons to provide students with a means of proving their knowledge of course topics \cite{affia2022integrating}. They also propose spacing out the hackathons through the period of the course instead of hosting one non required hackathon at the end of the course.
\end{itemize}
\subsection{Broader Ideas for Software Engineering Course Designs}

There were also broader ideas for software engineering courses that can be designed.
\begin{itemize}
\item \textbf{Robotic Systems.} One idea discussed was teaching more specific topics with a software engineering and design focus. Hildebrandt et al. developed a course for undergraduates on software development for robotics \cite{hildebrandt2022preparing}. Students learned about robotics and how to build systems with it, and they were also involved in focused sessions/assignments for debugging. They suggested that teaching more specific engineering topics such as robotics in a focused classroom setting, with an emphasis on software engineering. 
\item \textbf{Coding Camps.} Moster et al. created a summer coding camp for autistic high school students \cite{moster2022can}. In the course, students learned soft skills, such as kindness, teamwork, proactive communication, and scrum. These skills, which are often lacked by autistic students, are critical for learning through the remote learning environment. The researchers chose to used written and pre-recorded videos because students progressed through course at different paces.
\item \textbf{Capstone Courses.} Capstone projects are a useful learning complement to the learning environment because they allow students to understand the link between the concepts they learn in a software engineering class and how they can be utilized in real-world scenarios. However, as Bütt et al. explains, capstone projects remove the experience of risks experienced by the real-world \cite{butt2022student}. Thus, they suggest introducing an entrepreneurship aspect to capstone projects.
\end{itemize}

\section{Pressing Challenges to SE Research Community}
Through paper presentations, follow-up discussions, and Birds of a Feather sessions, several challenges in software engineering (SE) research became apparent.

Given the number of papers on SE for ML as well as ML for SE, there were several conversations on what ML robustness really means from a software engineering and testing point of view. Given the almost infinite space of inputs and the non-deterministic nature of ML models, testing machine learning systems is challenging, and it is hard to define the sufficiency of testing to achieve robustness.

Another broad issue and concern that was discussed was the issue of replicability. Many recognized that many published experimental artifacts are not easily reproducible. This is because of a wide range of reasons, including lack of effort in creating and documenting replication packages and lack of maintenance of these packages. PhD students mentioned that the incentive is often to publish more papers and to explore new research directions instead of spending time to create higher quality replication packages and to maintain them after graduation. Several papers also addressed this concern. Timperley et al. discussed the challenges that affect the quality of published artifacts and proposed suggestions to mitigate these \cite{timperley2021understanding}. Daoudi et al. investigated the reproducibility of published Android malware detection software and discussed how barriers to reproduction can be broken down \cite{daoudi2021lessons}. On a similar note, these challenges also arise for researchers/domain experts in various fields and data scientists \cite{epperson2022strategies} who may lack software engineering experience. Thus future research to understand how the software engineering community can help to bridge this gap is important.

Another broad topic discussed was around what is required for industry to adopt tools developed at ICSE 2022 and how academic research can have a larger impact in software engineering. Maybe software engineering researchers should have a foot in industry to better understand the risks and challenges of adopting new tools. On the other hand, maybe more software developers and project managers from industry settings should attend research conferences such as ICSE. Those working in industry could also host grand challenges or otherwise spread word of challenged that need to be overcome in practice. An issue with this though was that this could bias the research space to a subset of all the problems that need to be addressed. Maybe industry needs to be more willing to adopt new tools and technologies.

These challenges are important for SE researchers to consider in their future research, but from these discussions, it is also apparent that the solutions to these challenges are still unclear.

\section{Research from UW}

UW researchers contributed to ICSE 2022 in the areas of reliability and safety, SE communities, and mutation testing. Winston et al. investigated repairing machine-learning-based brain-computer interfaces with fault localization \cite{winston2022bci}, Liang et al. explored mining OSS skills from GitHub \cite{liang2022oss}, and Kaufman et al. presented work on prioritizing mutants in mutation testing \cite{kaufman2022prioritize}.

\bibliographystyle{ACM-Reference-Format}
\bibliography{paper}

\end{document}